\documentclass[final,5p,times,twocolumn]{elsarticle}

\usepackage{graphics}

\usepackage{amssymb}

\journal{Central European Journal of Physics}

\begin{document}

\begin{frontmatter}

\title{Spring-block model for a single-lane highway traffic}

\author[ubb]{Ferenc J\'arai-Szab\'o}
\author[ubb]{Bulcs\'u S\'andor}
\author[ubb]{Zolt\'an N\'eda}

\address[ubb]{Faculty of Physics, Babes-Bolyai University, 
              RO-400084 Cluj-Napoca, str. Kogalniceanu nr. 1, Romania}

\begin{abstract}
A simple one-dimensional spring-block chain with 
asymmetric interactions is considered to model an idealized single-lane 
highway traffic. The main elements of the system are blocks (modeling cars), 
springs with unidirectional interactions (modeling distance keeping 
interactions between neighbors), 
static and kinetic friction (modeling inertia of drivers and cars) and  
spatiotemporal disorder in the values of these friction forces 
(modeling differences in the driving attitudes). The traveling chain of cars correspond
to the dragged spring-block system. Our statistical analysis for the spring-block 
chain predicts a non-trivial and rich complex behavior. 
As a function of the disorder level in the system a dynamic phase-transition is observed. 
For low disorder levels uncorrelated slidings of blocks are revealed while for
high disorder levels correlated avalanches dominates.
\end{abstract}


\end{frontmatter}




\section{Introduction}

Road traffic is a non-linear complex phenomenon which strongly affects our
everyday life. Within this phenomenon various forms of 
collective behavior may be detected. The simplest possible form of 
traffic is the one on a single highway lane. 
The motion of the queue in this simple form of traffic is
primarily governed by the leading car and the statistics of driving attitudes. 
This simple situation becomes already
quite complex if the first car is moving slowly and the differences between the
driving attitudes are substantial. In such situations the queue will evolve
non-continuously in avalanches of different sizes, and jams of different magnitude
will continuously appear. Unfortunately, such situations are common in our
everyday life, so understanding and modeling it is important in order to optimize
our society.

Traffic studies tend to discover fundamental rules in different kind of 
transport systems that are essential for our social and economic efficiency. Accordingly, 
in this field many theoretical models have been developed. 
The study of traffic begins early in 1935 with the pioneering work of 
Greenshields \cite{Greenshields1935}. In 1955 Lighthill and Whitham 
published the oldest and most popular macroscopic model based on the theory of fluid 
dynamics \cite{Lighthill1955}. After these early works, and motivated by the 
explosive increase in road traffic, an avalanche of 
publications in leading international journals began. These models can be classified into four categories: 
microscopic models, macroscopic models, cellular automata models and non-traditional models. 
Detailed description and analysis of such traffic models can be found in the work of 
Darbha et al. \cite{Darbha2008}. More recently, the nonlinear effects of small perturbations have 
encouraged the development of stochastic models \cite{Mahnke2005} for the highway traffic.
A complete review of these traffic studies
can be found in the recent work of Nagatani \cite{Nagatani2002}, or 
in Kerner's book \cite{Kerner2004}. Despite of many studies in the field, 
due to the complexity of the problem, there are several phenomena which are still 
not well understood. 

\begin{figure}
\centering
\includegraphics[width=70mm]{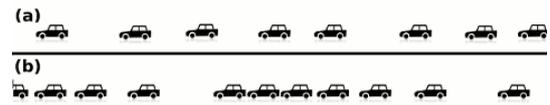}
\caption{Traffic with laminar flow structure (a) and traffic with jams (b).\label{fig1}}
\end{figure}

In order to get familiar with the complexity of the traffic and with the 
typical non-linear avalanche-like phenomena that can easily occur in highway traffic let us imagine
that we are driving on a highway in the middle of a car queue. The traffic is normal
with a laminar flow structure shown in Figure \ref{fig1}a. Suddenly, the driver of
the car located in front of us sees something and slows down a bit. 
To avoid collision, we should slow down only a bit, we should only reduce the 
foot pressure on the gas pedal. But, being concentrated to the car audio system 
instead we observe a little later the event and consequently we are forced to step the 
brakes. Seeing our braking lights, the driver behind us suddenly brakes and  
stops at the lane. A small event caused thus a substantial change in the traffic flow pattern. 
A wave consisting of cars with higher density forms and the laminar flow turns
into a start-stop-start-stop motion of cars caused by temporary traffic
jams that appears and propagate as sketched in Figure \ref{fig1}b. 

Such situations are investigated in the present work by means of computer simulations. 
A simple spring-block type model with asymmetric interactions is used to model the phenomenon  
and the effect of parameters having major influences on the dynamics of the system will be 
investigated. 
For some model parameters interesting collective behavior is found and it is 
statistically studied.
Up to our knowledge real experimental data describing the dynamics of
cars in such idealized situation is not available. Therefore, at
this stage the study has to limit itself on pure modeling, and the
test of the results rely simply on our everyday experience.

\section{The spring-block model}

A model family with broad interdisciplinary applications is the 
spring-block type models. Previously we have used them successfully to study complex systems 
which have shown self-organization phenomena. This model family was introduced in 1967 
by R. Burridge and L. Knopoff \cite{Burridge1967} to explain the empirical law of 
Guttenberg and Richter \cite{Gutenberg1956} on the size distribution of earthquakes. 

This earthquake model consists of simple elements: blocks that can slide with friction 
on a horizontal plane connected in a lattice-like topology by springs.
In the one-dimensional version of the model, the involved tectonic plates are modeled by 
two surfaces and their relative motion is governed by the sliding of the spring-block 
system on the lower surface. The upper surface (to which each block is connected 
by spring) is dragged with a constant velocity. Sliding of the blocks is realized by 
avalanches, as they are following the motion of the upper plate. These avalanches 
correspond to earthquakes and the energies dissipated by friction shows power law
type distribution as in case of the Guttenberg-Richter law.

Subsequently, due to the spectacular development of computers and computational methods, 
this simple model proved to be very useful in describing many phenomena in different 
areas of science. Most of the collective phenomena that occurs on mezoscopic scale in solid 
materials can be modeled by spring-block models. The model is almost always usable when 
one has to deal with avalanches of different nature, complex dynamics or structure 
formation by collective behavior. Recently, this model has been applied successfully to explain 
the formation of structures obtained in drying granular materials in contact with a 
surface \cite{Leung2000}, to study the fractures with curved topologies in 
wet granular materials \cite{Leung2001}, to understand the formation of 
self-organized nanostructures produced by capillary effects 
\cite{Jarai2005,Jarai2006,Jarai2007}, to study the magnetization processes and 
Barkhausen noise \cite{Kovacs2007} and to describe glass 
fragmentation \cite{Horvath2008}. Based on the previous examples, one can conclude 
that the spring-block type models are fully applicable to the study of complex
collective behavior in systems of different nature.

\begin{figure}
\centering
\includegraphics[width=70mm]{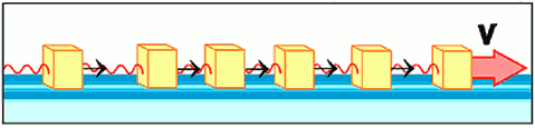}
\caption{Sketch of the spring-block chain considered here to model the idealized
single lane highway traffic.\label{fig2}}
\end{figure} 

The spring-block model for idealized single lane highway traffic can be
imagined as a chain of blocks (see Figure \ref{fig2}), modeling cars, connected 
by springs which represent some distance keeping interactions between them.
This interaction represents the will of drivers to keep a certain distance 
from the car ahead. Therefore, in order to be realistic, these springs 
cannot be real, bi-directional 
mechanical springs because in case of traffic without accidents this distance
keeping interaction acts only on the car in the back. The front car is never
pulled back or pushed ahead by the car from behind unless there is a collision. 
Accordingly, the action-reaction principle is violated and, in this sense,
this spring-block system cannot be a considered as a typical mechanical system.

Another important ingredient that has to be incorporated in the model is the 
disorder in the system generated
by the driving style of drivers. From our point of view, this should be
the drivers inertia which indicates how quickly the driver can react 
to a certain event or how quickly can follow the velocity change
of the car ahead. This is introduced via friction forces between
the blocks and the plane on which they slide. These friction forces 
are generated randomly and independently for each new position of 
each of the blocks. 
As a first approximation we consider
a normal distribution with a 
fixed mean $F_m$ and standard deviation $\sigma$. As a result of this, the disorder
introduced is both spatial and temporal. Spatial disorder 
means differences in driving styles while temporal disorder means
fluctuations of driving attitudes in time.

Moreover, in analogy with real mechanical systems static and kinetic 
friction forces are considered denoted by $F_s$ and $F_k$, respectively.
As in usual classical mechanics systems, the static force is considered
to be greater than the kinetic one, and for simplicity in our model their 
ratio is kept constant $f = F_k / F_s$. Here we have to note that for all
the presented simulations $f=0.8$ has been used. Of course, this parameter is 
adjustable one and its influence on the 
dynamics of the system is planned to be investigated later. 

Regarding the distribution of the friction forces one more comment needs to 
be added: when the static friction forces are randomly
updated in each new position of blocks the kinetic friction force is 
updated, too.
As a result, both the kinetic and static friction forces acting on blocks
will fluctuate in time and in space.  

The motion of the block queue is simulated in discrete simulation steps. For 
each step it corresponds the same time length $dt = 1$, fixing the unit 
for simulation time and the possibility to handle easily the 
stochasticity in the equation of motions. Due to this choice we do
not have possibility to define a velocity for the first block, and 
for us a given mean value of static friction $F_m$ and spring constant $k$ 
implicitly corresponds to a fixed drag velocity.
Accordingly, the motion of the queue is governed by the drag step 
(the movement of the first block) which is kept constant in time. 
In each unit of time the first block moves ahead in steps of length $d_0$.
In the present simulations
for the spring-block row a step limit of $d_{max} = 1$ is imposed.
This represents a kind of speed-limit for the units in the system.
It is important to realize that by changing the drag step value we do not
change the velocity of the first block. Instead of this we consider
either a finer (for small $d_0$ values) or a rough (for large $d_0$ values)
approximation of the continuous dynamics.

The length unit in the simulation is defined by the length of 
a single block $L=1$. In order to simulate the simplest dynamics without accidents, 
a minimum distance between blocks $d_{min} = 0.3$ is imposed. This distance 
is considered to be also the equilibrium length $l_0$ of the asymmetric springs.

The motion of each block is governed by the total force acting on it.
The force unit in the model has been selected by considering that all 
springs have the spring-constant $k = 1$. Unit elastic force acts on the block from 
behind when the distance between the two blocks is $(d_{min} + 1)$ expressed in 
simulation length units.

With this force unit we fix the mean value of the normal distribution for friction forces 
$F_m = 4$, and in the following the $\sigma$ parameter is considered to be 
the main parameter governing the disorder in the system.

Therefore, in this study, the model will have two free parameters: the
drag step $d_0$ of the first block and the disorder level in the static/kinetic 
friction $\sigma$. All other model parameters have been fixed as 
$d_{min} = 0.3$, $d_{max} = 1.0$ and $F_m = 4.0$ and they will not be 
explicitly noted in our later discussion.

Blocks are labeled after their ordinal number in the row so that the dragged block
(the first in the queue) has label 1, the next one label 2, and so on 
until the number of blocks in the queue $N$. Their position will be noted 
by $x^{(i)}$, where $i = \overline{1,N}$.

The simulation follows the typical steps of a simplified molecular dynamics 
simulation summarized bellow. 

(I) All blocks are 
visited and the spring force $F_{spring}$ acting on each block is calculated 
\begin{equation}
	F_{spring}^{(i)} (t) = k  \left[ x^{(i-1)}(t-1) - x^{(i)}(t-1) - L - l_0 \right]\,.
\end{equation}
If the block is at rest (its previous displacement is 0)
the spring force is compared to the static friction. 
If $F_{spring}^{(i)}(t) \leq F_s^{(i)}(t)$ the static friction 
equals the spring force and the total force acting on the block will be 
\begin{equation}
	F_t^{(i)} = 0\,,
\end{equation}
and the block will remain at rest in this step.
On the contrary, if $F_{spring}^{(i)}(t) < F_s^{(i)}(t)$ the block will start to move and the
kinetic friction force is added to the spring force and
\begin{equation}
	F_t^{(i)} = F_{spring}^{(i)} - F_k^{(i)}\,. \label{eq3}
\end{equation}  
The total force is similarly calculated, when the block is not at rest 
(its previous displacement differs from 0).

(II) All blocks are visited and, based on 
forces $F_t^{(i)}$ calculated in simulation step (1), 
their displacements $\Delta x^{(i)}$ are calculated. The displacement of the 
first block is $\Delta x^{(1)} = d_0$ which represents the constant drag step 
of the first block. For the rest of blocks $i = \overline{2,N}$
the displacement is calculated by using the equation of classical mechanics 
with $dt = 1$ 
\begin{equation}
	\Delta x^{(i)}(t) = \Delta x^{(i)}(t-1) + A \, F^{(i)}(t)\,, 
\end{equation} 
where $A = 1 / 2m$, $m$ being the mass of a block. In our simulations 
this constant is chosen to be $A = 1$ which fixes the mass of 
blocks to $m = 0.5$ expressed in simulation units.

(III) For the calculated displacements the block step 
limit $d_{max}$ is applied which means, that all 
displacement values grater than $d_{max}$ are set to be $d_{max}$.

(IV) All blocks are visited following their ordinal numbers and their new
positions are calculated
\begin{equation}
	x^{(i)}(t) = x^{(i)}(t-1) + \Delta x^{(i)}(t)\,.
\end{equation}
Here the only restriction is that the minimum distance $d_{min}$ between blocks
has to be respected.

(V) Quantities of interest for the selected block are detected and stored for 
later analysis.

Steps (I)-(V) are repeated until enough data is collected 
for the statistical analyses.

\section{Results of modeling}

\begin{figure}
\centering
\includegraphics[width=70mm]{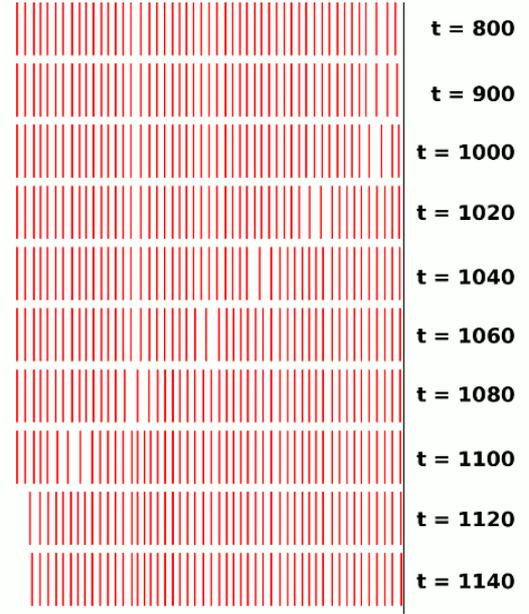}
\caption{Time evolution of a queue with 50 blocks.\label{fig3}}
\end{figure}

The dynamics of the system is illustrated with a short queue of 50 blocks 
in Figure \ref{fig3}. The first block is moving to right with small drag step
$d_0=0.01$. The snapshots are taken on the time steps 
printed on images. It has to be noted that between time steps 1000 and 1100 
the propagation of an avalanche through the whole system is observable.  

From the viewpoint of a single block in the row it's stop-time distribution
$g(\tau)$ is first measured. This stop-time distribution describes the distribution of time 
intervals during which a given block is not moving. In other words, this 
distribution function $g(\tau)$ determines the probability that the rest time of the 
block is $\tau$. Alternatively, the cumulative distribution function 
$g_>(\tau)$ may be defined, which gives us the probability that the stop-time of a block
is bigger than $\tau$. Based on the distribution function 
the mean stop-time and the stop-time standard deviation may be calculated.

\begin{figure}
\centering
\includegraphics[width=70mm]{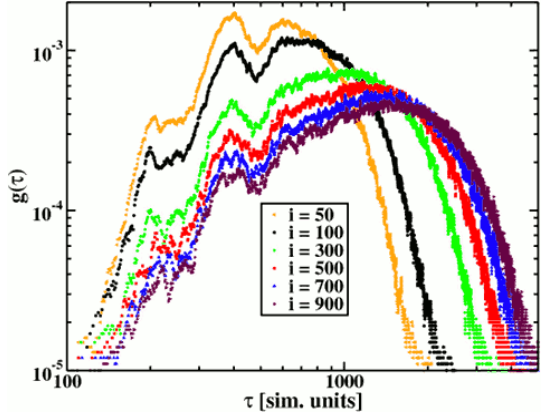}
\caption{Stop-time distributions for different block positions $i$. 
The plotted results are obtained from a statistics of 100\,000 stop-times.\label{fig4}}
\end{figure}

First, let us investigate how the position 
of a block in the queue is influencing its stop-time distribution. In Figure \ref{fig4}
the stop-time distributions of different blocks in the row are presented
for a fixed small drag step $d_0 = 0.01$ and disorder in the static friction 
forces $\sigma = 0.4$. 
In these simulations $N=1000$ blocks have been used and the
distribution function is constructed using the statistics of 100\,000 stop-times.  
From this results it can be learned that after a certain 
number of blocks (transient distance), the cumulative stop-time distribution 
converges to the same function. Therefore, in our later investigations a block 
after the transient distance has to be selected in order to be sure that the
average single-block behavior is studied. The form of the single block stop-time 
distribution function will be discussed later.

\begin{figure}
\centering
\includegraphics[width=70mm]{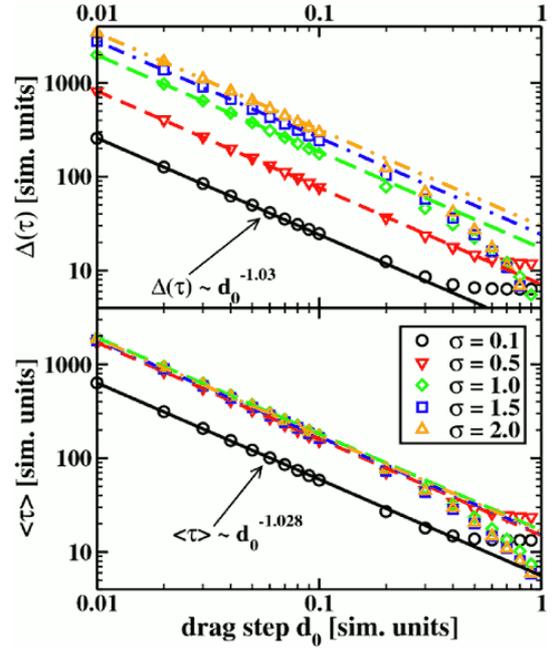}
\caption{Stop-time averages (bottom) and standard deviations (top) depending on 
         the drag step calculated for the $500$th block in a queue of $1000$ blocks.
         These quantities are calculated from 100\,000 stop-times. 
         \label{fig5}}
\end{figure}

Another more illuminating analyses may be performed if one looks at the stop-time averages and
standard deviations. First, their dependence on the drag step is investigated.
The position of the studied block is fixed at the middle of the queue $i = 500$ and 
several disorder levels $\sigma = 0.1, 0.5, 1.0, 1.5, 2.0$ are imposed. 
The results are plotted in Figure \ref{fig5}.
As one may observe, except for high drag step values the average stop-time $<\tau>$ 
and the standard deviation of stop-times $\Delta(\tau)$ are scaled with the drag step 
following a $1/d_0$ functional dependence.
In order to understand these scalings we have to remind ourself again, that
the drag step is not equivalent with a drag velocity, and all this is a result of the
fact that the simulation time-step was fixed a priori to $dt = 1$. This means that if 
one wants to get close to a realistic, continuum dynamics, the drag step has 
to be lowered to infinitesimally small values. Changing the drag step
is in fact equivalent to a change in the time-length $dt$ of a simulation step and 
thus influences only the real time of a simulation step, and implicitly everything 
else which is measured in time units. Therefore, this scaling suggests only
that the simulations are performed in the right continuum limit.
Based on our results we find that this limit is reached if drag steps smaller than 
$0.1$ are selected.

\begin{figure}
\centering
\includegraphics[width=70mm]{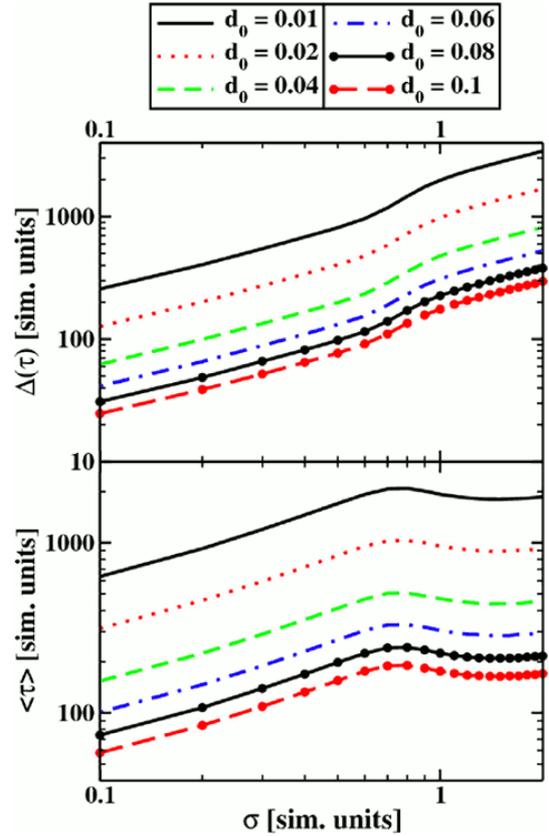}
\caption{Stop-time averages (bottom) and standard deviations (top) depending on 
         the disorder in static friction calculated for the $500$th block in a 
         queue of $1000$ blocks. These quantities are calculated from 100\,000 
         stop-times.\label{fig6}}
\end{figure}

Concerning the dynamics of the studied system interesting and non trivial results
may be obtained if one looks at the influence of the friction force 
disorder level $\sigma$ on the stop-time averages and standard deviations. In this sense
a series of simulations have been performed for several drag step values between 
$0.01$ and $0.1$. The position of the studied block is fixed at the middle of the 
queue ($i = 500$), and the averages and standard deviations are calculated 
for 100\,000 stop-times. The simulation results are plotted on Figure \ref{fig6}.

On one hand, as it is expected based on our previous considerations,
the statistical behavior of the system for different and small enough drag steps 
is the same. On the other hand, in case of all curves, at a certain disorder 
level ($\sigma \simeq 0.7$) a maximum in the stop-time average may be detected. 
This result shows us that in our simple traffic system there is a ''worst'' disorder 
level in driving attitudes (modeled by friction forces) for which the average 
stop-time of a block in the row is maximum. Moreover, in case of the stop-times 
standard deviation as a function of the mean disorder in the friction force value,
at this ''worst'' disorder level an inflection point may be detected. 

Therefore, based on these observations we can conclude that this simple 
spring-block system apparently exhibits two different types of behavior
as a function of the disorder level in the system. At low disorder levels
in the friction forces the average stop-time scales with the disorder level. 
This scaling law disappears however for higher disorder levels. Since this change
in the system dynamics appears for a certain disorder level, we consider
it as a \textit{disorder induced phase transition} similar with the one recently 
obtained in the 
spring-block model of magnetization phenomena \cite{Kovacs2007}. This phase transition is
a high order phase transition, because a peak only in the 
derivative of the stop-time's standard deviation is detectable.

\begin{figure}
\centering
\includegraphics[width=70mm]{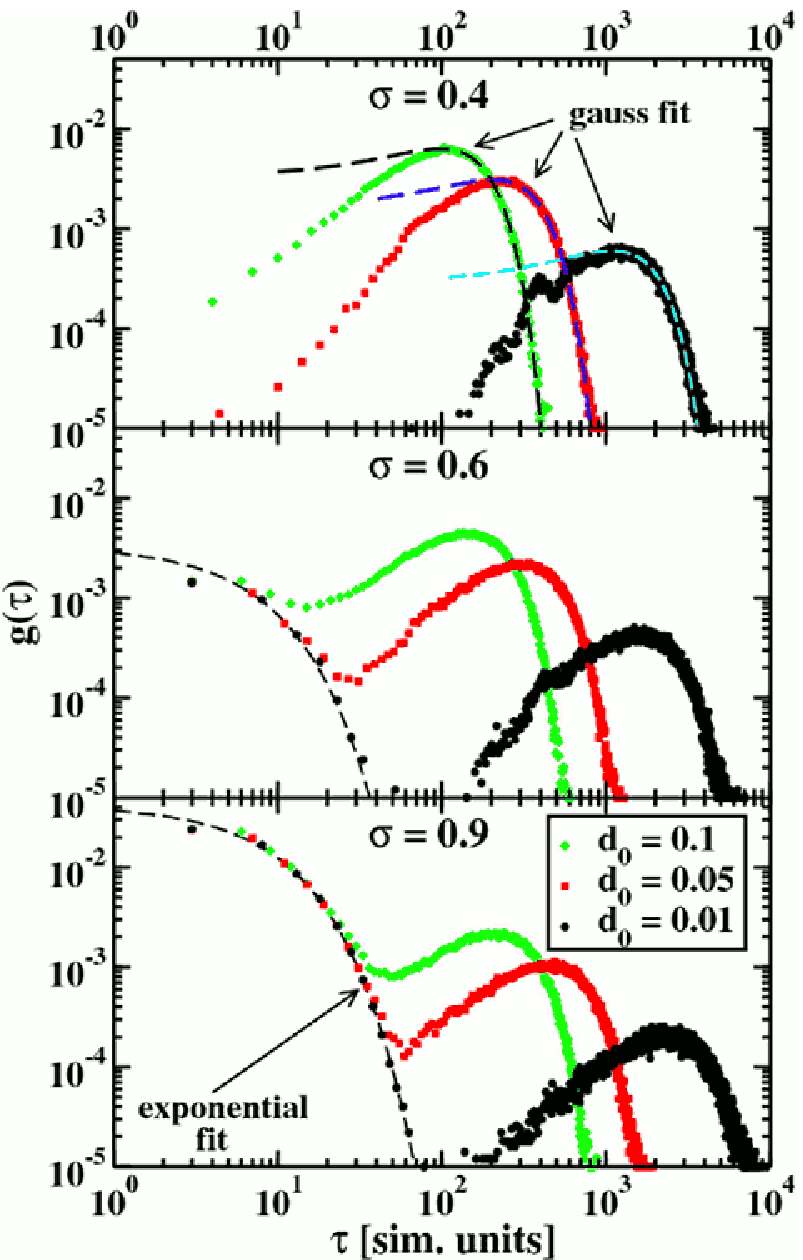}
\includegraphics[width=70mm]{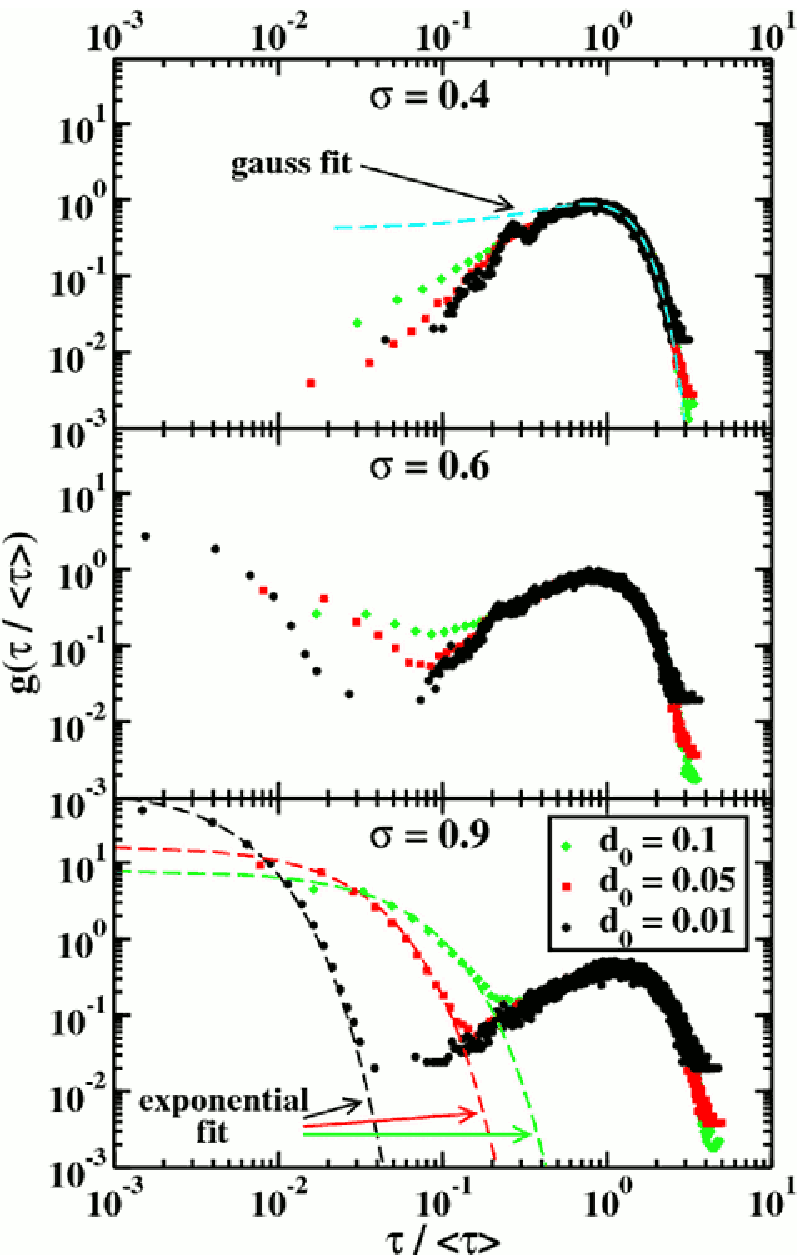}
\caption{Stop-time distributions (top panel) and normalized stop-time distributions (bottom panel) of a single block in the middle of a queue for different disorder levels of the friction force.\label{fig7}}
\end{figure}

In order to take a closer look at the nature of this athermal phase transition
detected in our spring-block system stop-time distributions of a single block are analyzed.
Three disorder levels $\sigma = 0.4, 0.6, 0.9$ are selected which correspond to values lower than,
close to, and higher than the critical disorder level. 

The distribution functions are constructed from 100\,000 simulated 
stop-times and they are plotted in the left panel of Figure \ref{fig7} for different drag steps of 
$d_0 = 0.01, 0.05, 0.1$. As it is immediately observable, near the critical disorder level 
(central graph on the left panel of Figure \ref{fig7}) there are
two peaks with the same magnitude in the stop-time distribution function best 
visible for drag step $d_0 = 0.01$. At lower disorder levels (top graph on left panel 
of Figure \ref{fig7}) the first peak disappears and the stop-time distribution of long 
stop-times will be close to a normal distribution. This may be explained if we suppose
that in this region the motion of blocks is independent to each other and it is mainly
influenced by friction forces generated randomly from a normal distribution. 
On the contrary, at higher disorder 
levels (bottom graph on left panel of Figure \ref{fig7}) the first peak, which has an 
exponential nature, will have the most important contribution to 
the stop-time distribution function suggesting stops that are induced by collective effects. 
It is extremely important to note that 
these exponential curves falls into a single one in case of different drag steps. 
Taking into account our previous results concerning the $<\tau> \sim 1/d_0$ scaling of 
average stop-times, this observation gives us an evidence that this part of the stop-time 
distribution is not depending on the stop-time averages and it is caused by an
avalanche-like collective motion of blocks.

The same data may be analyzed from another point of view if the distribution of 
normalized stop-times $\tau / <\tau>$ is constructed. This curves are plotted on the
right panel of Figure \ref{fig7}. It is immediately observable that the 
Gaussian parts of distribution functions fall into a single curve. This confirms 
our previous presumption that the block motions at small disorder levels are mainly 
independent to each other which causes that the resultant distribution of normalized 
stop-times to be the same. 

Therefore, our simulation results show that the studied spring-block system may exhibit
non-trivial, critical behavior. Through the detected disorder induced higher order phase 
transition the independent block motions will be organized to a highly correlated 
avalanche-like collective motion of the queue.
 
\section{Conclusions}

In summary, in the present paper a simple spring-block type model with asymmetric interactions 
has been used to model the simplest possible, idealized form of traffic which happens on 
one single highway lane. The spring-block chain dragged by constant steps of the 
leading block predicts non-trivial, complex behavior even for this simple form of traffic.
Our investigations have been performed by a series of computer simulations which targeted 
the analysis of the effect of parameters on the model dynamics. As a result, it was concluded
that at low disorder levels in friction forces the dynamics is governed by uncorrelated sliding
of the blocks. On the contrary, at higher disorder levels the dynamics of the block sliding
self-organizes into an avalanche-like motion of blocks. The 
transition between these domains is realized through a disorder induced phase transition
whose effects are detectable in stop-time averages and standard deviations, too.

\section*{Acknowledgments}
This work was supported by CNCSIS-UEFISCSU, project number PN II-RU PD\_404/2010.

\bibliographystyle{elsarticle-num}

\end{document}